\pdfoutput=1
\documentclass[10pt,twocolumn,letterpaper]{article}

\pdfoutput=1

\usepackage
[
        letterpaper,
        left=.75in,
        right=.75in,
        top=.75in,
        bottom=.75in,
]{geometry}
\usepackage{authblk}

\usepackage{times}  
\usepackage{helvet} 
\usepackage{courier}  
\usepackage[hyphens]{url}  
\usepackage{graphicx} 
\usepackage{subfig}
\usepackage{amssymb}
\usepackage{amsmath}
\usepackage{tabularx}
\usepackage[eulergreek]{sansmath}
\usepackage{pdfpages}
\usepackage{graphicx}
\usepackage{tabularx}
\usepackage{enumitem}
\usepackage{dblfloatfix}
\usepackage{booktabs}
\usepackage{multirow}
\usepackage{ulem}
\usepackage{tikz}
\usepackage{pgfplots}
\usepackage{caption} 
\usepackage{pgfplotstable}
\usepgfplotslibrary{fillbetween}
\usepackage{xcolor}
\usepackage{tabularx}
\usepackage{booktabs}

\definecolor{color0}{rgb}{0.12156862745098,0.466666666666667,0.705882352941177}
\definecolor{color1}{rgb}{0.866666666666667,0.517647058823529,0.32156862745098}
\definecolor{color2}{rgb}{0.298039215686275,0.447058823529412,0.690196078431373}
\definecolor{color3}{rgb}{0.333333333333333,0.658823529411765,0.407843137254902}
\definecolor{color4}{rgb}{0.768627450980392,0.305882352941176,0.32156862745098}
\definecolor{color5}{rgb}{0.505882352941176,0.447058823529412,0.701960784313725}
\definecolor{color6}{rgb}{0.576470588235294,0.470588235294118,0.376470588235294}
\definecolor{color7}{rgb}{0.854901960784314,0.545098039215686,0.764705882352941}
\definecolor{color8}{rgb}{0.8,0.725490196078431,0.454901960784314}
\definecolor{color9}{rgb}{0.392156862745098,0.709803921568627,0.803921568627451}
\definecolor{burnt_orange}{HTML}{a72207}
\definecolor{burnt_blue}{HTML}{3f4f73}
\colorlet{background_grey}{white!70.9019607843137!black}

\usetikzlibrary{pgfplots.groupplots}
\usetikzlibrary{shapes.geometric}
\usetikzlibrary{positioning,fit,shapes.geometric,backgrounds, calc}
\usetikzlibrary{arrows,decorations.markings}
\usetikzlibrary{shapes.arrows}
\usetikzlibrary{patterns}
\tikzset{textnode/.style={inner sep=0pt,outer sep=0,execute at begin node={\strut}}}
\tikzstyle{state} = [textnode,circle, draw, inner sep=0pt, outer sep=0]
\usepgfplotslibrary{groupplots}
                    
\pgfplotsset{every axis/.append style={
                    xlabel={$x$},          
                    ylabel={$y$},          
                    label style={font=\sffamily\scriptsize},
                    tick label style={font=\sffamily\scriptsize},
                    xticklabel style = {font=\sffamily\scriptsize},
                    title style = {font=\footnotesize\sffamily},
                    ylabel near ticks,
                    y label style={font=\sffamily\scriptsize},
                    xlabel near ticks,
                    x label style={font=\sffamily\scriptsize},
                    legend cell align={left},
                    legend style={draw=none, font=\sffamily\scriptsize},
                    },
                    legend image code/.code={
                    \draw[mark repeat=2,mark phase=2]
                        plot coordinates {
                        (0cm,0cm)
                        (0.15cm,0cm)        
                        (0.3cm,0cm)         
                        };%
                    }
                    }
\pgfplotsset{compat=newest}                    
                    
\def\addlegendimage{\csname pgfplots@addlegendimage\endcsname}

\newcommand{\ignore}[1]{}
\newcommand{\nop}[1]{}
\newcommand*{\eg}{\textit{e.g.}}

\newcommand*{\ie}{\textit{i.e.}}

\newcommand{\tw}[1]{\textcolor{red}{TW: #1}}
\newcommand{\rk}[1]{\textcolor{violet}{RK: #1}}
\newcommand{\tf}[1]{\textcolor{blue}{TF: #1}}

\let\showplots\iftrue 

\begin{document}
\title{Competition Dynamics in the Meme Ecosystem}
\author{Trenton Ford, Rachel Krohn, and Tim Weninger \\ Department of Computer Science and Engineering \\ University of Notre Dame
\\ \{tford5, rkrohn, tweninger\}@nd.edu}

\date{}

\maketitle

\begin{abstract}
The creation and sharing of memes is a common modality of online social interactions. The goal of the present work is to better understand the collective dynamics of memes in this accelerating and competitive environment. By taking an ecological perspective and tracking the meme-text from 352 popular memes over the entirety of Reddit, we are able to show that the frequency of memes has scaled almost exactly with the total amount of content created over the past decade. This means that as more data is posted, an equal proportion of memes are posted. One consequence of limited human attention in the face of a growing number of memes is that the diversity of these memes has decreased at the community level, albeit slightly, in the same period. Another consequence is that the average lifespan of a meme has decreased dramatically, which is further evidence of an increase in competition and a decreasing collective attention span.
\end{abstract}

\section{Introduction}

With the rise of social media platforms, the cost historically associated with producing and consuming information has decreased to unprecedented levels; users and organizations can easily share their thoughts, stories, and others' content with diverse and widespread audiences with very little effort. Due to the ease of production, the volume of content produced has increased to the point that any individual user can only see a small portion of what is available. The reduction in content production cost and increase in availability has induced a change in scarcity dynamics: from content scarcity to consumer scarcity~\cite{terranova2012attention}. This shift in scarcity has birthed new research areas to help users see relevant content -- such as recommender systems -- as part of the broader \textit{attention economy}~\cite{shapiro1998information}.

\begin{figure}
    \centering
    \includegraphics[width=0.43\textwidth]{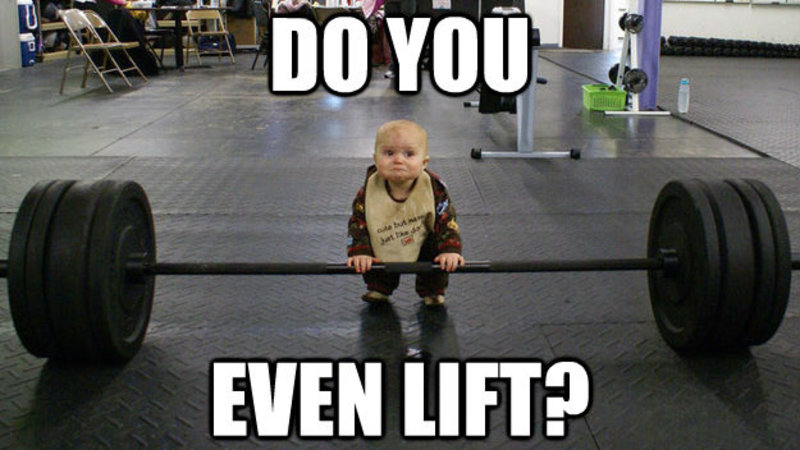}
    \caption{Meme image with text meant to be condescending to the subject. The text of this meme and others like it are frequently used without the image in humorous or sarcastic contexts.}
    \label{fig:doyoulift}
\end{figure}

The attention economy seeks to explain the allocation of cognitive resources in the creation and consumption of information. Though this concept existed long before the advent of social media~\cite{simon1969designing}, recent work has focused on how this model governs the dynamics of content consumers and curators in the socio-digital space~\cite{glenski2017consumers,glenski2018guessthekarma}. The main focus has been on the consumption of information~\cite{weng2012competition,wu2007novelty,hodas2012visibility}, but others focus on the production and curation of information~\cite{ciampaglia2015production,leskovec2009meme,glenski2017rating,simmons2011memes,huberman2009crowdsourcing}. One of the primary questions at the center of online social media is this: how does the limited attention of users shape the information landscape?

One particularly compelling subset of the information landscape is the production and resharing of compelling memes, which are short phrases and images. For example, the image in Figure~\ref{fig:doyoulift} is a meme with text that is meant to be condescending to the subject; oftentimes, the text of the meme is independent of its image and is written in plaintext in comments and tweets. The dynamics of these viral messages are not well understood despite widespread attempts to predict and simulate their spread or popularity~\cite{blythe2019massive,coscia2013competition}. Yet this narrow subset of the information landscape is an increasingly visible and influential communication mode with exciting properties. A meme's popularity can be quantified by how many times it is reproduced or shared, how long it stays relevant, and how many times it is mutated -- in the case of meme images.

Intuition about how memes are created, transmitted, consumed, or mutated are often derived from their association with genes and the process of gene evolution~\cite{dawkins2016selfish} and more recently, memes have been considered through the lens of epidemiology and disease transmission~\cite{wang2011epidemiological,kubo2007possibility}. Indeed, its etymology is a portmanteau of mind+gene, which begs the question: rather than continuing the economic analogy, are memes better situated in the realm of \textit{ecology}? And if so, what kind of understanding can be gleaned from this perspective?
    
In the present work, we derive findings about competition and diffusion of information from the ecological perspective. There are many compelling examples that motivate this perspective. Foremost is the concept of \textit{competition}, which is a driving force in genealogical, economic, epidemiological, and ecology fields~\cite{nissenbaum2017internet}. In analogical terms, the goal of competing memes is their continued existence within the minds and communication patterns of people. \textit{Survival}, therefore, follows as a natural extension of competition, which presumes that memes are designed with survival in mind~\cite{wiggins2015memes,leskovec2009meme}.
    
The differences between the ecological perspective and others are nuanced. Fundamentally, each perspective offers a unique interpretation of information dynamics. For example, within the economics perspective, human behavior (\eg, attention) is the primary focus, and memes just one of many possible factors. From an epidemiological perspective, memes are treated as a contagion (\eg, a virus), but epidemiological models typically do not consider landscapes with multiple viruses and their interactions. The genealogical perspective treats memes as genes and explores gene-gene interactions, but the gene perspective does not natively consider gene-environment interactions.

In taking the ecological perspective, we consider a meme to be a single species existing within the same environment or habitat. The ecological perspective shifts the focus away from the human users and back to the memes and the environments they exist within -- wherein memes seek both longevity and a large population, competing for limited environmental resources -- human attention.

Within the perspective of the \textit{meme ecology}, we ask the following research questions:
\begin{itemize}[leftmargin=.5in]
    \item [RQ1:] How does the collective user attention scale? Do more users permit a larger or smaller number of memes?
    \item [RQ2:] How do memes compete for attention? How does the introduction of a new meme impact the ecosystem of existing memes?
    \item [RQ3:] How have the dynamics of collective attention changed over time?
\end{itemize}

In summary, by using well-known metrics and concepts from ecology, we perform an ecological analysis of the dynamics of text-memes on Reddit. The results of this analysis and the behavior they suggest are compelling and strongly support the case for the ecology of memes. We find that memes comprise a relatively constant fraction of all activity on the platform, even as social media increases in popularity. This suggests that as more memes are created their lifecycle duration becomes shorter, which further suggests that the collective human attention span on social media is decreasing. 

Although the current work focuses on short, frequently repeated texts, \ie, memes, we further hypothesize that our findings are likely to apply to a number of other communication modalities like image-memes and hashtags.

\begin{table}[t]
    \centering
    \small{
    \begin{tabular}{@{}rrl@{}}
        \toprule
        \textbf{Tokens} & \textbf{Count} & \textbf{Examples} \\
        \midrule
        1 & 69  & thicc; yeet; wat; mfw; impossibru \\
        2 & 80  & moms spaghetti; zerg rush; y tho \\
        3 & 69  & winter is coming; u wot m8 \\
        4 & 48  & do you even lift; kill it with fire \\
        5 & 25  & hello darkness my old friend \\
        6 & 25  & shrek is love shrek is life \\
        7 & 18  & still a better love story than twilight \\
        8 & 18  & this is why we cant have nice things\\
        \bottomrule
    \end{tabular}
    }
    \caption{Meme dataset consists of 352 text memes, ranging in length from 1 to 8 tokens. Some memes reference current pop-culture events, while others seem unconnected to trends of the time.}
    \label{tab:memetable}
\end{table}

\begin{figure}[t]
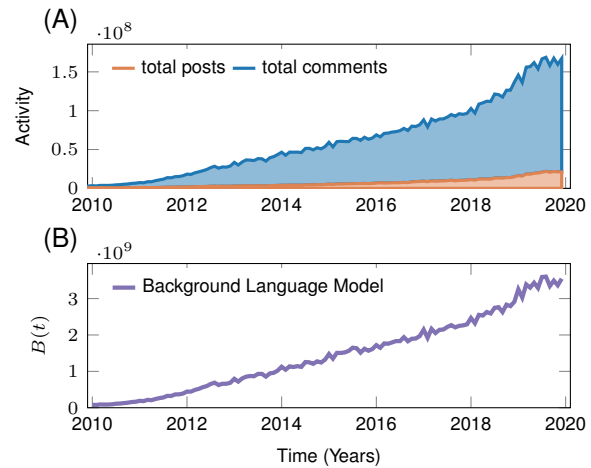

    \centering
    \include{plots/reddit_growth}
    \caption{(A) Stacked line plot representation of Reddit contributions between 2010 and 2020. The lower (orange) region shows the number of posts per month; the upper (blue) region shows the number of comments per month. (B) The unigram background model is used to compute normalized meme-frequencies. This background behavior closely mirrors the growth of Reddit, but is one order of magnitude larger.}
    \label{fig:reddit_growth}
\end{figure}

\section{Data Collection}
Using a comprehensive dataset of the 352 most popular memes from KnowYourMeme.com, we identified their individual occurrences on Reddit. The memes were selected from the Confirmed category on KnowYourMeme, and include text-based memes that ranged from 250 thousand to 13 million page views each. Note that the tracking of rapidly-evolving image templates is outside the scope of the present work; therefore, image-memes are not included in this analysis. Extended meme text (\eg, \textit{copypasta}) is truncated to include only the 8-token prefix. The final set contains meme-phrases that range in length from 1 to 8 word tokens as shown in Table~\ref{tab:memetable}.  Additionally, we collected all posts and comments from Jan. 2010 to Jan. 2020; the number of monthly posts and comments is plotted in Figure \ref{fig:reddit_growth}.

The questions raised in the present work are considered human subjects research, and relevant ethical considerations are present. We sought and received research approval from the Institution Review Board of \textit{redacted}.

\section{Collective Attention to Memes is Stationary}
Previous work has shown that innovation and technological development is accelerating. Moore's Law is one example of this phenomenon where a compounding increase in circuit density has led to remarkable increases in computational power~\cite{schaller1997moore}; similar effects have been shown in genome sequencing~\cite{mardis2011decade} and telecommunications bandwidth~\cite{eldering1999there}. In online social systems, the early empirical evidence suggests that a similar pattern exists: that social innovations are accelerating~\cite{lorenz2019accelerating,rosa2003social,hutchins2011acceleration,rosa2013social}.  

This is the basis for \textbf{RQ1}: How does collective user attention of memes scale? Does the presence of larger groups result in super-scaling effects like those found in population densities~\cite{pan2013urban} and software development~\cite{thomas2019dynamics} where collections of individuals produce more than the sum of their parts? 

At first glance, Figure~\ref{fig:reddit_growth} appears to show that our data supports these claims: more posts, comments, and memes are being made at an accelerating pace year over year. But how much attention is paid to individual memes? To answer this question, we first need to measure collective attention.

\subsection{Measuring Collective Attention}
Because we cannot collect the number of users who viewed or thought about a particular meme, instead, we estimate the collective attention for each meme by the number of times it appears, \ie, its frequency. So our first task is to extract the daily frequency $F_m(t)$ of each meme, such that $F_m(t)$ is the frequency of meme $m$ on day $t$.
Given the scale variance of social media sites, simply counting tokens would be insufficient to fully determine whether memes are truly increasing in prevalence or if their frequency increases mirror the increase in content that Reddit has experienced overall. In order to produce as accurate assessment of the meme ecology, we need to carefully normalize the frequency of meme occurrences.

To do this, we constructed a set of 5000 randomly selected words from Reddit to serve as a background language model. 
Then, for each day, we count the number of occurrences of each token in the background set, such that $B(t)$ gives the background sum of all 5000 words on day $t$. As the number of posts and comments on Reddit grows over time, so too do the occurrences of the background set. As seen in Figure~\ref{fig:reddit_growth}, the background growth closely mirrors the growth of Reddit.

Likewise, we expect the number of meme occurrences to increase at a similar rate. In order to control for the growth of Reddit over time we compute the \textit{normalized meme-frequency} for each meme $\hat{F}_m(t) = F_m(t) / B(t)$.


\begin{figure}[t]
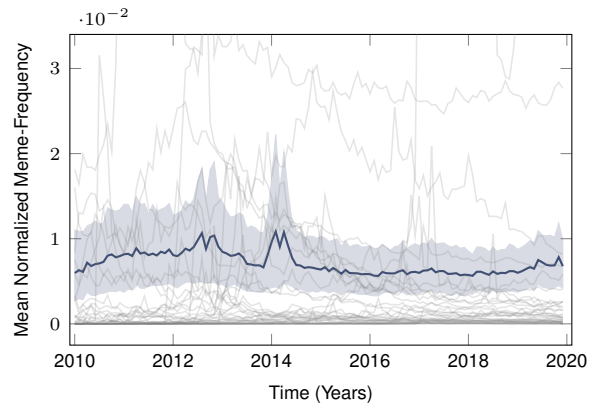

    \centering
    \include{plots/normalized_aggregate_meme_freq}
    \caption{Average normalized meme-frequency from 2010--2020 and 95\% confidence interval (shaded region). Light grey lines show the individual normalized meme-frequency for a random 10\% sample of individual memes. Overall, meme occurrence has remained consistent over the past decade (Pearson $R=+0.03$, $p$-value$<0.01$).}
    \label{fig:agg_meme_freq}
\end{figure}

Figure~\ref{fig:agg_meme_freq} illustrates the mean average normalized meme-frequency along with its 95\% confidence interval from 2010 to 2020. A selection of individual memes are also plotted in light grey. We find that the occurrence of memes remains remarkably consistent when controlled for Reddit's overall activity, even as the occurrence of individual memes varies widely. Correlation analysis finds almost no association between time and the normalized meme-frequency (Pearson $R=+0.03$, $p$-value$<$0.01).

Competition for limited attention is not a new concept in socio-digital media studies. However, much of the previous analysis has focused on network-level simulation or mathematical modeling as an attempt to predict the behavior of large social systems~\cite{weng2012competition,gleeson2014competition}. User-level attention analysis seeks to understand the impact of competition and content volume on individual actors~\cite{hodas2012visibility}. Other work attempts to determine which features will make a meme successful within a competitive environment~\cite{coscia2013competition,lakkaraju2013s}. These works are valuable extensions of the classic attention economy, yet fail to demonstrate the system-wide effect of competition. By showing that meme activity accounts for a stable fraction of all Reddit content, we have demonstrated the ability for competition to act as a kind-of global bandwidth cap. Even with a rapidly expanding user base, and a seemingly insatiable hunger for fresh content, Reddit appears only to sustain a relatively limited population of memes.

\section{Competition Among Memes}

The prevalence of individual memes rises and falls over time. Popularity is fleeting, but as a meme dies out, another always seems to rise to take its place. The ebb and flow of what is or is no longer popular has long been studied as the \textit{diffusion of innovations} ~\cite{gleeson2014competition}. The study of the production and evolution of innovations (\eg, new ideas, behaviors, memes, and other information patterns) has recently become particularly compelling because of the availability of vast amounts of human behavior happening through online social platforms.

The dynamic behavior of memes, in particular, suggests that these ideas exist in constant competition with one another. The competition among memes is not unlike competition observed in markets, where innovations eventually replace outdated products, or ecological systems, where competition among individuals in a group exerts a selective pressure that rewards certain genetic innovations.

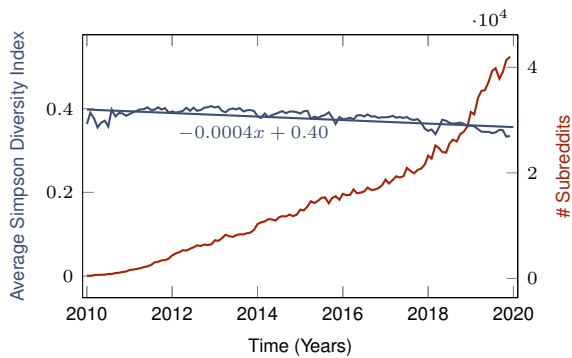
\begin{figure}[t]
    \centering
    \pgfplotstableread{
x_data	year	avg_diversity	std_dev_diversity	subreddit_count	95_confidence   subreddit_percentage
0	2010.00	0.363189263	0.015432615	514	0.00133730935	0.012229651
1	2010.08	0.390516603	0.015900815	504	0.00139154884	0.01199172
2	2010.17	0.376204634	0.014389544	611	0.00114323911	0.014537581
3	2010.25	0.35516892	0.013804154	671	0.00104636180	0.015965167
4	2010.33	0.366195301	0.014143815	685	0.00106105653	0.01629827
5	2010.42	0.371316194	0.013552852	713	0.00099648994	0.016964477
6	2010.50	0.357342645	0.012525493	845	0.00084574184	0.020105165
7	2010.58	0.397916711	0.012275136	845	0.00082883731	0.020105165
8	2010.67	0.381454246	0.011345911	1006	0.00070195912	0.023935854
9	2010.75	0.389626484	0.010885387	1061	0.00065573815	0.025244474
10	2010.83	0.388839652	0.01039059	1207	0.00058677383	0.028718266
11	2010.92	0.383053827	0.009792649	1313	0.00053017193	0.031240334
12	2011.00	0.387139381	0.008687045	1593	0.00042691617	0.037902401
13	2011.08	0.391350971	0.008760965	1660	0.00042175764	0.039496538
14	2011.17	0.395198547	0.008294864	1796	0.00038388148	0.042732399
15	2011.25	0.39834281	0.008011907	1934	0.00035729569	0.046015846
16	2011.33	0.398831271	0.007725877	2127	0.00032851822	0.050607914
17	2011.42	0.403044642	0.007459535	2288	0.00030581708	0.054438602
18	2011.50	0.397191949	0.007057182	2522	0.00027555952	0.060006186
19	2011.58	0.397441127	0.006371522	3008	0.00022778609	0.07156963
20	2011.67	0.403370055	0.00645075	3155	0.00022517768	0.075067215
21	2011.75	0.398012932	0.006093098	3417	0.00020437024	0.081301006
22	2011.83	0.40149429	0.00601575	3460	0.00020051728	0.08232411
23	2011.92	0.390246136	0.005831354	3768	0.00018625229	0.089652383
24	2012.00	0.393647262	0.00537577	4368	0.00015946591	0.10392824
25	2012.08	0.390661796	0.005203662	4730	0.00014833300	0.112541341
26	2012.17	0.392344012	0.005055871	4980	0.00014045428	0.118489614
27	2012.25	0.394688512	0.004871236	5351	0.00013054734	0.127316853
28	2012.33	0.402223589	0.004888799	5269	0.00013203406	0.125365819
29	2012.42	0.394845614	0.004742169	5632	0.00012387600	0.134002712
30	2012.50	0.402367711	0.004660684	5883	0.00011912083	0.139974779
31	2012.58	0.395088388	0.004591556	6256	0.00011380037	0.148849604
32	2012.67	0.396940871	0.004595954	6157	0.00011482186	0.146494087
33	2012.75	0.402239978	0.004520482	6416	0.00011063247	0.152656499
34	2012.83	0.403643711	0.004510741	6322	0.00011121207	0.150419948
35	2012.92	0.406100315	0.004459363	6469	0.00010868850	0.153917533
36	2013.00	0.402904626	0.004216348	7229	0.00009721161	0.172000286
37	2013.08	0.405251186	0.004264319	7112	0.00009912332	0.169216493
38	2013.17	0.394574086	0.00414818	7647	0.00009298843	0.181945799
39	2013.25	0.39137152	0.003968733	8267	0.00008556370	0.196697518
40	2013.33	0.3980817	0.004043404	8004	0.00008859462	0.190439934
41	2013.42	0.400580607	0.004081602	7875	0.00009016131	0.187370625
42	2013.50	0.39825692	0.004013481	8194	0.00008691315	0.194960622
43	2013.58	0.401551932	0.003927514	8359	0.00008420766	0.198886483
44	2013.67	0.393503259	0.003940911	8343	0.00008457591	0.198505794
45	2013.75	0.39606544	0.003929282	8555	0.00008327463	0.20354993
46	2013.83	0.397052543	0.003917577	8668	0.00008248345	0.20623855
47	2013.92	0.394107952	0.003770033	9228	0.00007693013	0.219562683
48	2014.00	0.381913127	0.003524785	10306	0.00006805917	0.24521164
49	2014.08	0.377907958	0.00349664	10626	0.00006649111	0.25282543
50	2014.17	0.388660935	0.003446775	10814	0.00006497054	0.257298532
51	2014.25	0.379733669	0.003370173	11216	0.00006237753	0.266863356
52	2014.33	0.384574345	0.003386798	11193	0.00006274963	0.266316115
53	2014.42	0.387611607	0.003481248	10958	0.00006518765	0.260724738
54	2014.50	0.391000322	0.003411137	11569	0.00006216483	0.275262319
55	2014.58	0.394187931	0.00339005	11800	0.00006117272	0.280758524
56	2014.67	0.389219264	0.003317745	11749	0.00005999781	0.279545076
57	2014.75	0.393155843	0.003318536	12102	0.00005913022	0.287944039
58	2014.83	0.393328421	0.003346282	11806	0.00006036759	0.280901282
59	2014.92	0.391648098	0.0033106	12083	0.00005903518	0.28749197
60	2015.00	0.388492794	0.003218918	13025	0.00005528527	0.309905066
61	2015.08	0.393585717	0.003202204	12873	0.00005532203	0.306288515
62	2015.17	0.395087554	0.00311793	13620	0.00005236783	0.324061957
63	2015.25	0.375906711	0.003036593	14624	0.00004921954	0.347950225
64	2015.33	0.380896199	0.002999334	14274	0.00004920813	0.339622642
65	2015.42	0.378051048	0.002977335	14726	0.00004809159	0.350377121
66	2015.50	0.382364919	0.002996381	15264	0.00004753848	0.363177806
67	2015.58	0.384948769	0.002933477	15353	0.00004640538	0.365295391
68	2015.67	0.391314339	0.003070905	14254	0.00005041769	0.33914678
69	2015.75	0.379788194	0.002978766	15254	0.00004727452	0.362939875
70	2015.83	0.363490702	0.002945255	15563	0.00004627625	0.370291941
71	2015.92	0.380164156	0.00297303	14913	0.00004771996	0.354826429
72	2016.00	0.375197492	0.002881226	16036	0.00004459747	0.381546075
73	2016.08	0.375796984	0.002906275	15765	0.00004537025	0.375098147
74	2016.17	0.37847258	0.002923479	15822	0.00004555652	0.376454353
75	2016.25	0.374242263	0.00282624	16857	0.00004266759	0.401080207
76	2016.33	0.384054791	0.002864731	16139	0.00004420040	0.383996764
77	2016.42	0.384411969	0.002869931	16267	0.00004410605	0.38704228
78	2016.50	0.381362226	0.002838343	16483	0.00004333379	0.392181589
79	2016.58	0.386395676	0.002780212	17214	0.00004153513	0.409574342
80	2016.67	0.381028641	0.00281156	16754	0.00004257626	0.398629518
81	2016.75	0.381991421	0.002818346	16846	0.00004256231	0.400818482
82	2016.83	0.378008548	0.002775811	17081	0.00004163055	0.40640986
83	2016.92	0.379423595	0.00274836	17654	0.00004054431	0.420043303
84	2017.00	0.379523081	0.002643278	18730	0.00003785735	0.445644674
85	2017.08	0.380026096	0.002728166	17987	0.00003987207	0.427966404
86	2017.17	0.383106473	0.002636544	18671	0.00003782051	0.444240881
87	2017.25	0.378379715	0.0026292	19347	0.00003705033	0.460325014
88	2017.33	0.380276937	0.002651703	19119	0.00003758962	0.454900188
89	2017.42	0.377857472	0.002573888	19358	0.00003626057	0.460586738
90	2017.50	0.380183216	0.002518968	20947	0.00003411418	0.498393966
91	2017.58	0.374714797	0.002546835	20402	0.00003494928	0.485426729
92	2017.67	0.372977893	0.002571445	19911	0.00003571948	0.473744319
93	2017.75	0.375258045	0.002569987	20541	0.00003514745	0.488733969
94	2017.83	0.367483128	0.002524836	20791	0.00003432170	0.494682243
95	2017.92	0.352267568	0.002491946	21574	0.00003325414	0.513312237
96	2018.00	0.346915283	0.002374389	23273	0.00003050678	0.553736706
97	2018.08	0.350377647	0.002433557	22719	0.00003164595	0.540555331
98	2018.17	0.339082824	0.002319431	25189	0.00002864475	0.599324276
99	2018.25	0.354642014	0.002357128	24710	0.00002939112	0.587927383
100	2018.33	0.373735105	0.002379643	23960	0.00003013272	0.570082562
101	2018.42	0.371157492	0.002421737	23836	0.00003074542	0.567132218
102	2018.50	0.36609047	0.002318267	25578	0.00002841181	0.60857979
103	2018.58	0.364270692	0.002267898	26294	0.00002741343	0.625615646
104	2018.67	0.365360397	0.002285197	25952	0.00002780396	0.617478408
105	2018.75	0.36681031	0.002225816	27337	0.00002638646	0.650431845
106	2018.83	0.364006754	0.002184489	27760	0.00002569846	0.660496324
107	2018.92	0.358196099	0.002163865	28784	0.00002499890	0.684860453
108	2019.00	0.359801712	0.002066861	31530	0.00002281467	0.750196293
109	2019.08	0.357717135	0.002099742	31084	0.00002334331	0.739584573
110	2019.17	0.350974877	0.001979929	34254	0.00002096802	0.815008684
111	2019.25	0.345176729	0.001948895	35555	0.00002025820	0.845963501
112	2019.33	0.344619037	0.001932106	35675	0.00002004989	0.848818673
113	2019.42	0.344661784	0.001883345	37303	0.00001911262	0.887553832
114	2019.50	0.341545122	0.001808933	39319	0.00001788062	0.935520712
115	2019.58	0.343457677	0.001796764	39829	0.00001764626	0.94765519
116	2019.67	0.34923154	0.00187162	37860	0.00001885339	0.900806586
117	2019.75	0.348583718	0.001840693	39107	0.00001824381	0.930476576
118	2019.83	0.333219422	0.001784913	41340	0.00001720650	0.983606557
119	2019.92	0.334756689	0.001781894	42029	0.00001703602	1

}{\data}

\begin{tikzpicture}

\begin{axis}[%
    width=2.9in,
    height=2.0in, 
    %
    xlabel={Time (Years)},
    xticklabels={ 2010, 2012, 2010, 2012, 2014, 2016, 2018, 2020},
    ylabel={\textcolor{burnt_orange}{\# Subreddits}},
    scaled y ticks=true,
    axis y line*=right,
    xticklabel/.append style={
        /pgf/number format/.cd,%
        scaled x ticks = false,
        set thousands separator={},
        fixed
        },
    xmin=2009.9,
    xmax=2020.1,  
]
\addplot [color=burnt_orange, thick, solid] table [x=year, y=subreddit_count] {\data};
\end{axis}

\begin{axis} [
    width=2.9in,
    height=2.0in, 
    xticklabel/.append style={
        /pgf/number format/.cd,%
        scaled x ticks = false,
        set thousands separator={},
        fixed
        },
    xmin=2009.9,
    xmax=2020.1,    
    xlabel near ticks,
    xticklabels={},
    ymin=-0.051796550427371, 
    ymax=.57693942341841,
    scaled y ticks=false,
    ylabel={\textcolor{burnt_blue}{Average Simpson Diversity Index}},
    xlabel={},
    ytick pos=left,
]

\addplot [color=burnt_blue, solid, thick] table [x=year, y=avg_diversity] {\data}; 
\addplot [name path=upper,draw=none] table[x=year ,y expr=\thisrow{avg_diversity}+\thisrow{95_confidence}] {\data};
\addplot [name path=lower,draw=none] table[x=year
,y expr=\thisrow{avg_diversity}-\thisrow{95_confidence}] {\data};
\addplot [fill=burnt_blue!20] fill between[of=upper and lower];
\addplot[forget plot, thick, burnt_blue, mark=none, domain=2010:2020] {-0.0042*x + 8.84}
node[below=0mm, pos = .39] {\scriptsize{\textcolor{burnt_blue}{$-0.0004x+0.40$}}};
\end{axis}

\end{tikzpicture}
    \caption{Average Simpson's Diversity Index across subreddits, and number of subreddits, over time. Meme diversity on Reddit is decreasing (Pearson $R=-0.63$, $p$-value$<0.01$) with a small slope (0.04\% per month) despite an increase in the number of subreddits containing memes.}
    \label{fig:avg_diversity}
\end{figure}

\begin{table*}[t]
    \centering
             \footnotesize{{
    \begin{tabular}{@{}llllll@{}}
        \toprule
        \textbf{\small{\textrm{Rank}}} & \textbf{\small{\textrm{2011}}} & \textbf{\small{\textrm{2013}}} & \textbf{\small{\textrm{2015}}} & \textbf{\small{\textrm{2017}}} & \textbf{\small{\textrm{2019}}} \\
        \midrule
        1 & \textcolor{color2}{/r/pics} & \textcolor{color1}{/r/funny} & /r/4chan & /r/me\_irl & /r/aww \\
        2 & \textcolor{color3}{/r/AskReddit} & \textcolor{color3}{/r/AskReddit} & \textcolor{color6}{/r/WTF} & /r/aww & /r/ComedyCemetery \\
        3 & \textcolor{color1}{/r/funny} & \textcolor{color6}{/r/WTF} & \textcolor{color5}{/r/gaming} & \textcolor{color2}{/r/pics} & /r/NBA2K \\
        4 & \textcolor{color5}{/r/gaming} & \textcolor{color2}{/r/pics} & /r/TumblrInAction & /r/stevenuniverse & /r/Right\_Wing\_Politics \\
        5 & \textcolor{color0}{/r/reddit.com} & /r/videos & /r/Smite & \textcolor{color3}{/r/AskReddit} & \textcolor{color5}{/r/gaming} \\
        6 & \textcolor{color4}{/r/politics} & /r/AdviceAnimals & \textcolor{color2}{/r/pics} & /r/worldnews & \textcolor{color1}{/r/funny} \\
        7 & \textcolor{color6}{/r/WTF} & /r/trees & \textcolor{color3}{/r/AskReddit} & /r/woahdude & /r/madlads \\
        8 & /r/comics & \textcolor{color5}{/r/gaming} & \textcolor{color1}{/r/funny} & /r/nba & /r/TheNewsFeed \\
        9 & /r/IAmA & /r/4chan & /r/dogecoin & /r/Drama & /r/wow \\
        10 & /r/fffffffuuuuuuuuuuuu & /r/IAmA & /r/sex & /r/TwoBestFriendsPlay & /r/OutOfTheLoop \\
        \bottomrule
    \end{tabular}
    }}
    \caption{Top 10 most innovative subreddits by year. Colored subreddit names show the top 10 most innovative subreddits from 2010 to 2020 in aggregate.}
    \label{tab:innovative_subreddits}
\end{table*}

Continuing the ecological analogy, one way to assess the health of an ecological system is to examine the \textit{diversity} of the species living within the system. From this perspective, the ecology of a social media environment ought to have several different memes simultaneously appearing in abundance. This leads us to \textbf{RQ2}: How do memes compete for attention? Specifically, as Reddit grows and new memes are introduced, does the overall diversity increase or decrease?

One way to measure the diversity of an ecological system is with Simpson's Diversity Index ($D$)~\cite{gregorius2008generalized}. This index takes into account both the number of species present, as well as the population of each species. It is formally defined as:

\begin{equation}
    D = 1 - \frac{\sum_{i=1}^{R} n_i(n_i-1)}{N(N-1)}
\end{equation}

\noindent Where $R$ is the total number of species in a community (numbered $1$ through $R$), $N$ is the total number of organisms, and $n_i$ is the number of individuals belonging to species $i$. For our purposes, we consider a subreddit to be a community, each meme as its own species, and each meme occurrence as a species entity. $D$ ranges from 0 to 1, where 0 indicates no diversity and 1 indicates infinite diversity. 

We compute the Simpson's Diversity Index for each subreddit in each month. We then compute the average monthly Simpson Diversity across all subreddits, resulting in a monthly average diversity of Reddit at the community level. Figure \ref{fig:avg_diversity} shows the average diversity and 95\% confidence intervals over time. Overall, the diversity of Reddit communities appears to be decreasing at a small (0.48\% per year) but steady rate (Pearson $R=-0.63$, $p$-value$<$0.001). 

This indicates that subreddit communities, on average, are using slightly less diverse meme subsets. This finding, coupled with the growth in the number of subreddits using memes, suggests a slow Balkanization of  Reddit communities. That is to say, as more subreddit communities come into existence, human attention -- and the memes which that attention supports -- moves towards more niche subreddits, which aligns with findings from other subreddit analysis~\cite{marchal2020polarizing}. 

\subsection{Innovative Communities}

Ecological research focuses on the environments in which species exist; the communities and biomes that spawn \textit{new} species are often the focus of study because they can host exotic species along with high levels of species diversity~\cite{macarthur1965patterns}. The same is true in social media, where the study of the creation and diffusion of innovations within social environments is a central topic of research. The detection and analysis of highly innovative communities are of great interest in this line of work~\cite{kolleck2013social}. As a direct consequence of our findings in RQ2 we further ask: Are there certain communities that consistently introduce new memes? These \textit{meme-nurseries} are the places where many new linguistic and cultural innovations are born and cultivated. We further ask: do these innovative communities persist over time, or is their high-innovation only temporary?

To answer these questions, we first define a meme \textit{entry event} as the first time that a meme is used within the Reddit ecosystem. The subreddit in which it first appears is defined as the landing space or \textit{beachhead} for the new meme. From an ecological perspective, a new meme is analogous to new species, and subreddits are the host habitats. Because Reddit itself exists within the even larger environment of the social internet, tracing the true spread of a meme between specific communities is intractable. For this reason, it is necessary to consider that a single meme may have multiple beachheads within Reddit. 

Under these circumstances, we elect to rank subreddits based on the order in which a meme first appears; this methodology reasonably controls for the uncertainty about which community was first, second, etc., to articulate the innovation. We constrict our analysis to the first 1000 subreddits to use each meme. Based on this ordering, we compute the mean reciprocal rank (MRR) of each community with respect to each of our 352 memes. These MRR values consider the rank for each subreddit for each year in the dataset. The top 10 most innovative subreddits in the odd-numbered years (due to space constraints) are given in Table \ref{tab:innovative_subreddits}

It is clear that certain subreddits contribute a disproportionate number of new memes to Reddit. The colors in Table~\ref{tab:innovative_subreddits} represent the top 10 subreddits from all 10 years combined. It is unclear \textit{why} these subreddits are more innovative than others, but there may be conditions within these particular communities that increase their ability to produce viral content more regularly than others. No matter the cause, optimal meme-nursery conditions appear to be transient, as subreddits can be highly innovative one year, and not the next.

There are a few conclusions to be drawn here. Early in Reddit's history, massive and highly contributive subreddits -- like /r/reddit.com and /r/AskReddit -- were the primary beachheads for new memes. However, as years progressed the set of top contributing subreddits became less consistent. Each new year comes with new meme beachheads. 

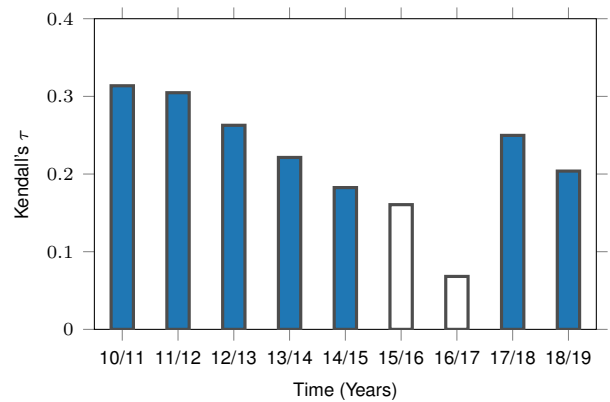
\begin{figure}[t!]
    \centering
    \begin{tikzpicture}

\begin{axis}[
height=2.25in,
width=3.25in,
tick align=outside,
xlabel={Time (Years)},
xtick={0,1,2,3,4,5,6,7,8},
xticklabels={ 10/11, 11/12, 12/13, 13/14, 14/15, 15/16, 16/17, 17/18, 18/19},
xmin=-0.5, xmax=8.5,
ylabel={Kendall's $\tau$},
ymin=0, ymax=0.40,
]
\draw[draw=white!30!black,fill=color0,very thick] (axis cs:-0.2,0) rectangle (axis cs:0.2,0.313661202185792);
\draw[draw=white!30!black,fill=color0,very thick] (axis cs:0.8,0) rectangle (axis cs:1.2,0.304710793082886);
\draw[draw=white!30!black,fill=color0,very thick] (axis cs:1.8,0) rectangle (axis cs:2.2,0.262761853895911);
\draw[draw=white!30!black,fill=color0,very thick] (axis cs:2.8,0) rectangle (axis cs:3.2,0.22135465245392);
\draw[draw=white!30!black,fill=color0,very thick] (axis cs:3.8,0) rectangle (axis cs:4.2,0.182563293060531);
\draw[draw=white!30!black,fill=color0,very thick] (axis cs:6.8,0) rectangle (axis cs:7.2,0.24976289942772);
\draw[draw=white!30!black,fill=color0,very thick] (axis cs:7.8,0) rectangle (axis cs:8.2,0.203800510064147);

\draw[draw=white!30!black,very thick] (axis cs:4.8,0) rectangle (axis cs:5.2,0.160575858250277);
\draw[draw=white!30!black,very thick] (axis cs:5.8,0) rectangle (axis cs:6.2,0.0681647940074906);

\end{axis}

\end{tikzpicture}
    \caption{Kendall's rank correlation coefficient ($\tau$) of subreddit innovation rankings for pairs of consecutive years. Solid bars represent $p<0.01$ and hollow bars represent $p>0.05$. Subreddits that consistently use new memes before other communities are ranked higher, but rankings change each year. A higher $\tau$ means more correlation between year-pairs, while lower indicates more change in their relative rankings. In general, subreddit beachhead rankings have become less stable, indicating greater turnover in the top subreddits.}
    \label{fig:meme_rank_shift}
\end{figure}

We quantify the degree of change in subreddit ranks by computing Kendall's ($\tau$) coefficient between consecutive years. Larger $\tau$ values indicate more similarity, smaller $\tau$ indicates less similarity, and a negative $\tau$ represents dissimilarity. Figure \ref{fig:meme_rank_shift} illustrates $\tau$ for each pair of years where solid bars represent statistical significance $p<0.01$ and hollow bars vice versa $p>0.05$; there were no $p$-values between 0.01 and 0.05. Until the 2017/2018 pairing the rank correlation trended downward, indicating increased turnover in the topmost innovative subreddits.

The 2017 to 2018 evaluation showed a return to high-rank correlation, indicating less change in the top-ranked subreddits. There are a few potential explanations for this behavior. First, this may be due to a major Reddit policy change: beginning in June 2017, Reddit removed the default subreddits, which included many /r/pics, /r/funny, and many of the other most innovative subreddits, and instead introduced /r/popular, which was a mix of posts from various subreddits as a means to expose new users to a wider variety of communities\footnote{\url{https://www.reddit.com/r/announcements/comments/6eh6ga/reddits_new_signup_experience/}}. This change essentially means that users ``subscribe'' to a wider variety of subreddits by default, providing a greater opportunity for innovations from niche subreddits to become more easily accessed. Another potential explanation for this trend reversal is due to the fact that the meme set used for our analysis is biased towards more popular, and therefore older, memes. Memes created during the last years of our analysis window are less likely to have become popular enough to appear in our top-memes dataset. This may result in fewer meme entries in more recent years.

\section{Changing Dynamics of the Meme Ecosystem}

Now that we have established some of the consequences of the competition of memes in a social media ecosystem, we turn our attention towards the dynamics of collective attention. Existing recent work suggests that these dynamics are accelerating~\cite{lorenz2019accelerating}, that is, new concepts are becoming viral faster and stay viral for a shorter duration. Instead of focusing on specific cultural artifacts like memes, the previous work on \textit{general} collective behavior focused on hashtags on Twitter, comments on Reddit, and n-grams in books, etc. Does this acceleration hold true for memes?  This is the basis for \textbf{RQ3}: How have the dynamics of collective attention on memes changed over time? Are we cycling through memes faster than we were a few years ago?

While we find this to be true in some ways, it is not true when the growth of the community is accounted for. 
In other words, relative to the number of words produced on Reddit, the number of memes is not changing. But what about the dynamics of individual memes? How have they changed over time? 

\subsection{Investigating Meme Dynamics}

\begin{figure}[t!]
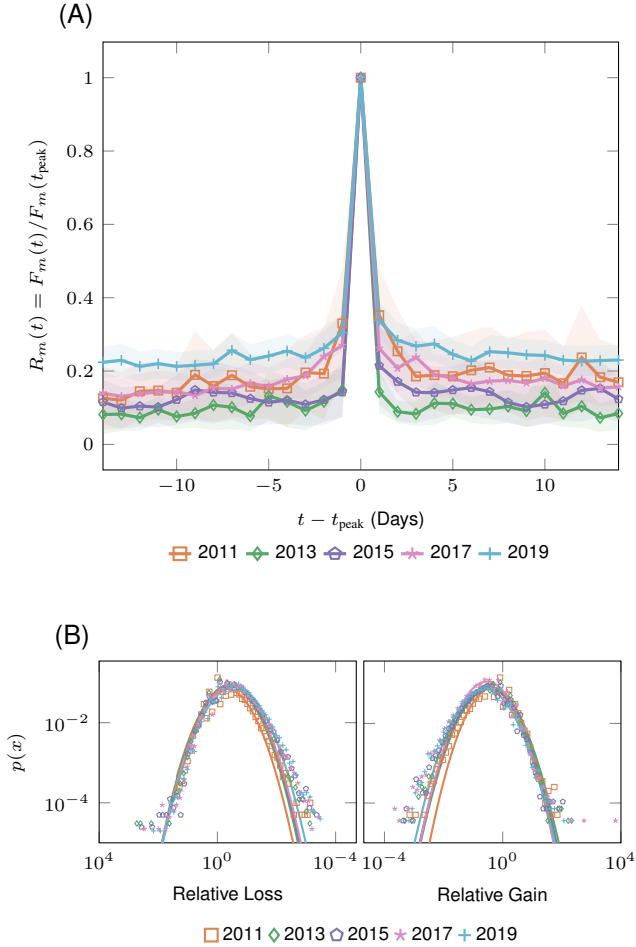

    \centering
    \include{plots/meme_peak}
    \vspace{-.3cm}
    \include{plots/gains}
    \caption{(A) Average relative meme-frequency, time-shifted so that the maximum frequency occurs on day 0. Shaded areas indicate 95\% confidence intervals. The width of the primary peak has not changed, suggesting that memes have not experienced significant acceleration. (B) Probability distribution of relative meme velocities divided into gains (right) and losses (left). Points give true distribution values, lines are fitted log-normal distributions. While small magnitude gains and losses have shifted slightly, larger velocities do not change. These stable velocities again indicate that memes have not accelerated.}
    \label{fig:meme_peak}
\end{figure}


Lorenz-Spreen et al. \cite{lorenz2019accelerating} used a variety of methods to analyze collective dynamics in the online social sphere. Here, we apply their methodology to our meme dataset.

\begin{figure*}[t]
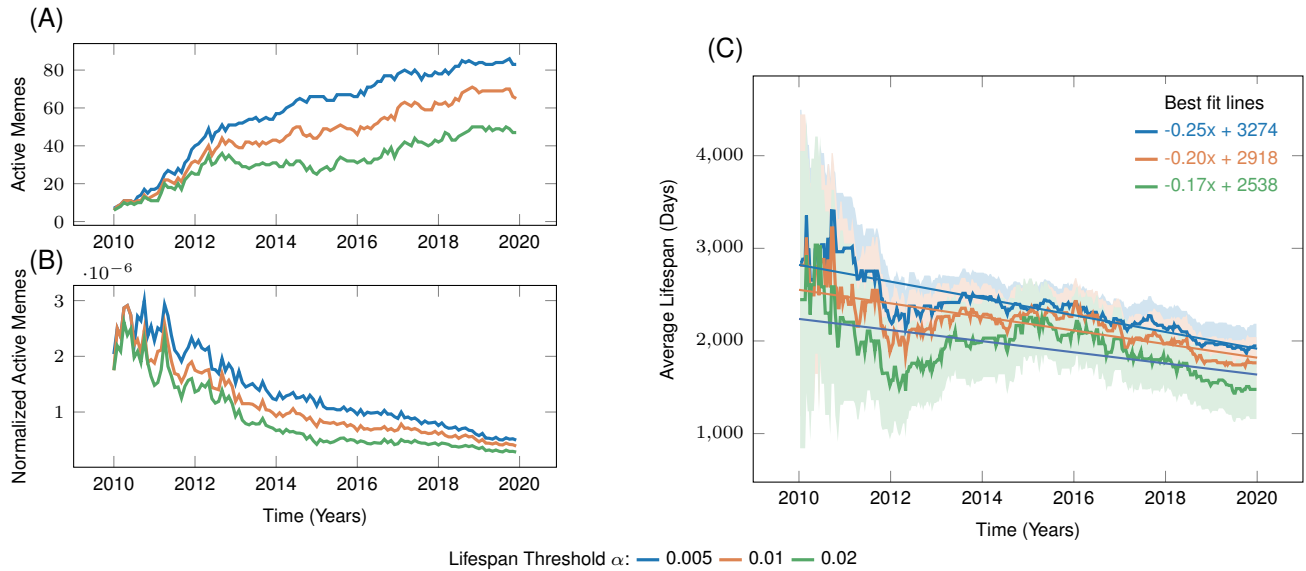

    \centering
    \hfill
    \begin{minipage}{.48\linewidth}
    \include{plots/active_memes}
    \end{minipage}
    \begin{minipage}{.48\linewidth}
    \include{plots/average_lifespan}
    \end{minipage}
    \hfill
    \caption{Results of lifespan analysis on meme set for $\alpha = 0.005, 0.01$, and $0.02$. Lifespan starts on the first day with a frequency $\geq \alpha \hat{F}_m(t_{\textrm{peak}})$, and ends on the last day with frequency $\geq \alpha \hat{F}_m(t_{\textrm{peak}})$, such that all days in the lifespan have a frequency $\geq \alpha$ times the maximum normalized meme-frequency. (A) Number of active memes per month, where a meme is active only during its defined lifespan. Overall, the raw number of active memes has increased. The dip in later years is likely the result of bias toward older memes in our dataset. (B) Number of active memes per month, normalized by total Reddit contributions. Reddit growth is outpacing the number of active text memes. (C) Average meme lifespan (in days) over time and corresponding best fit lines. Shaded area indicates 95\% confidence interval. Over the course of 10 years, the average lifespan has decreased ($\alpha = 0.005$: Pearson $R = -0.85$, $p$-value $<$0.01; $\alpha = 0.01$: Pearson $R = -0.77$, $p$-value $<$0.01; $\alpha = 0.02$: Pearson $R = -0.59$, $p$-value $<$0.01), indicating that memes are cycling faster.}
    \label{fig:active_memes_count}
\end{figure*}

First, we focus on the peaks of memes on Reddit to assess the pace of collective attention. For each meme, we compute its frequency across all of Reddit each day, such that $F_m(t)$ gives the frequency of meme $m$ on day $t$. We also identify the \textit{peak frequency} for each meme $F_i(t_{\textrm{peak}})$ and when this peak occurred. To ensure all memes are on the same scale, the frequencies of each meme are then normalized by that meme's peak to get a \textit{relative meme-frequency}: $R_m(t) = F_m(t) / F_m(t_{\textrm{peak}})$. In Figure~\ref{fig:meme_peak}(A), we illustrate the average relative peak frequency $R_m(t)$ for all memes in our dataset and group by the year of each meme's peak. Overall, there appears to be no change in peak dynamics over time. The difference between the peak and the baseline frequency (\ie, frequency before and after the peak) remains relatively stable, nor does it change a statistically significant amount from year to year. Furthermore, the changes seen do not show a trend over time. This suggests that memes have not exhibited a significant acceleration over the past decade.

Next, we look closer at the \textit{velocities} of memes. For each meme we compute relative gains $[\Delta F_m^{(g)} / F_m](t) = (F_m(t) - F_m(t-1) / F_m(t-1)$ and relative losses $[\Delta F_m^{(l)} / F_m](t) = (F_m(t) - F_m(t+1) / F_m(t+1)$, where gains are $>0$ and losses are $<0$. We analyze the distributions of losses and gains of all memes at all times, grouped by year. Both distributions fit well to a log-normal distribution. Gains and losses are shown in Figure \ref{fig:meme_peak}(B). While we observe some shift in gains and losses with small magnitudes, the larger velocities do not change significantly or regularly across years. 

Taken together, both of these analyses indicate that once the growth of Reddit is controlled for, the collective dynamics of Reddit memes have not accelerated. Rather, meme dynamics have remained remarkably consistent, even surrounding its peak.

\subsection{Meme Lifespans are Shrinking}

The previous analysis raises another interesting question about the collective dynamics of memes: Are meme lifespans growing or shrinking?

To answer this question we define the lifespan of a meme as follows. For a meme with a peak frequency of $F_m(t_{\textrm{peak}})$, the lifespan begins on the first day $u$ where $\hat{F}_m(u) \geq \alpha \hat{F}_m(t_{\textrm{peak}})$. Recall that $\hat{F}_m(t)$ is the normalized meme-frequency and is computed as $\hat{F}_m(t) = F_m(t) / B(t)$. The lifespan ends on the last day $v$ where the meme experiences a normalized frequency $\hat{F}_m(v) \geq \alpha \hat{F}_m(t_{\textrm{peak}})$, such that all days between the beginning, peak, and the end are continuous and above $\alpha \hat{F}_m(t_{\textrm{peak}})$. The threshold value $\alpha$ is very small; here we present results for $\alpha = 0.005, 0.01$, and $0.02$. Other values produced similar results. 

By defining the lifespan this way, each meme's lifespan captures the majority of its occurrence, but does not include very early, late, or anomalous uses. The lifespan is also determined using normalized meme-frequencies to control for Reddit growth. We compute the lifespan length in number of days for each meme. For a threshold of $\alpha = 0.005$, these lifespans range from a high of 4140 days to a low of 1 day. (For completeness, we include the full history of Reddit beginning in 2005 when defining lifespans.)

Based on this definition, we consider a meme as \textit{active} on all days contained within its lifespan. On any single day within our analysis window, there may be many active memes. Figure~\ref{fig:active_memes_count}(A) illustrates the number of memes active within each month. As expected, the number of active memes has increased over time. 

To more accurately assess the number of active memes, it is important to control for Reddit's growth. Figure \ref{fig:active_memes_count}(B) shows the normalized count of active memes, \ie, the number of active memes per month divided by $B(t)$. With this normalization applied, it is apparent that over time, the \textit{relative} number of active memes is decreasing. This decrease in the normalized number of active memes occurs despite our finding that normalized meme-frequency has remained stable (Figure \ref{fig:agg_meme_freq}). The most likely explanation for these contrasting findings is that the lifespan of memes is decreasing.

We can test this hypothesis directly by computing how lifespans have changed over time. 
Figure \ref{fig:active_memes_count}(C) plots the average meme lifespan and 95\% confidence interval in days for the active memes within each month. For example, if there were 40 memes active in June 2012 and the mean-average lifespan of these 40 memes was 2,400 days, then we would plot 2,400 for June 2012 (and the confidence interval). The lifespan is heavily dependent on the $\alpha$ threshold. So, we repeat these calculations for $\alpha = 0.005, 0.01$, and $0.02$. In all cases, we found that the average lifespan decreased over time ($\alpha = 0.005$: Pearson $R = -0.85$, $p$-value $<$0.01; $\alpha = 0.01$: Pearson $R = -0.77$, $p$-value $<$0.01; $\alpha = 0.02$: Pearson $R = -0.59$, $p$-value $<$0.01). Simply put, memes are rising and falling more quickly now than in the past.

To recap: the relative fraction of Reddit focused on memes has not changed over time (Figure \ref{fig:agg_meme_freq}), the meme diversity of subreddits has decreased (Figure \ref{fig:avg_diversity}), and the relative number of active memes has declined (Figure \ref{fig:active_memes_count}(B)). These results indicate that the production and consumption of new memes are accelerating. There is simply not enough attention to go around. As fast as new content is created, old content must be forgotten. Given the increasing volume of content on Reddit every day, the necessary consequence is that the collective attention span of Reddit is decreasing. 

An exception to this rule can be found in long-lasting memes, many of which live beyond the end of our data analysis window. For example, the meme `cool story bro' is still active on the last day of 2019, following an extended lifespan beginning in 2009; this is true for all $\alpha$ values tested. These long-lasting memes have transcended their origins and evolved from memes into slang, maintaining a consistent presence even as Reddit grows, and other memes come and go. However, despite their seemingly `undying' status, these durable memes must still compete for attention within the broader ecosystem.

In a content-consumption system driven by user attention, memes are forced into competition with one another for scarce attention. As collective attention decreases, memes appear to rise and fall at an accelerated rate. In a system that favors the newest, freshest content, no meme is immortal.





\ignore{
A reasonable question then becomes: if active memes \tf{these are not decreasing though} and meme lifespans are both decreasing, but the overall fraction of meme activity remains constant, where has all the meme activity gone? To reconcile this seemingly contradictory result, we next investigate how the maximum frequency of memes has changed.

\tw{^^this paragraph needs to conclude and then I think we end this section.}
\rk{not before that paragraph?}
\tw{perhaps... I'd like to put a slightly bigger bow on it though.}

\tw{I'm not sure how maximum meme frequency answers the contradiction above.}
\tf{I'm actually not convinced that there is a contradiction above.}

\tw{i'm not confident that this following paragraph is particularly interesting or informative}
\tf{rachel's lower comment may be pre-asking the same question.}

For each meme, we find its maximum frequency $F_i(t_{peak}$, and normalize it by the background set to get the normalized maximum meme frequency. For each month in our analysis window, we compute the average normalized maximum meme frequency for all memes whose maximum frequency occurred in that month. This gives us the average maximum frequency for each month - an aggregate measure of the maximum attention given to any one meme at that time. We visualize this quantity in Figure \ref{fig:avg_meme_peak}. As with the total attention paid to memes, the maximum attention is also stable when normalized by Reddit's growth. This further supports the notion of a limited attention economy: even in an ever-expanding social network, memes can only command so much user interaction.

\rk{having just written that - this max freq/peak analysis (last 3 paragraphs) bit may fit better in section 1 than here - since it actually has nothing to do with lifespans, and very little to do with meme dynamics. thoughts?}

\begin{figure}
    \centering
    \pgfplotstableread{
month	year	date	peaks_this_month	avg_peak_freq	std_dev_peak_freq	95_confidence
1	2010	2010	0	0	0	nan
2	2010	2010.083333	0	0	0	nan
3	2010	2010.166667	0	0	0	nan
4	2010	2010.25	0	0	0	nan
5	2010	2010.333333	0	0	0	nan
6	2010	2010.416667	0	0	0	nan
7	2010	2010.5	0	0	0	nan
8	2010	2010.583333	0	0	0	nan
9	2010	2010.666667	0	0	0	nan
10	2010	2010.75	1	6.41E-06	6.41E-06	nan
11	2010	2010.833333	0	0	0	nan
12	2010	2010.916667	1	8.19E-08	8.19E-08	nan
1	2011	2011	0	0	0	nan
2	2011	2011.083333	0	0	0	nan
3	2011	2011.166667	0	0	0	nan
4	2011	2011.25	0	0	0	nan
5	2011	2011.333333	1	2.24E-07	2.24E-07	nan
6	2011	2011.416667	2	3.48E-07	3.48E-07	3.1283E-06
7	2011	2011.5	1	4.15E-07	4.15E-07	nan
8	2011	2011.583333	2	1.08E-07	1.08E-07	9.70035E-07
9	2011	2011.666667	1	2.54E-06	2.54E-06	nan
10	2011	2011.75	2	4.22E-06	4.22E-06	3.79253E-05
11	2011	2011.833333	2	1.71E-08	1.71E-08	1.53844E-07
12	2011	2011.916667	7	4.20E-06	4.20E-06	3.88069E-06
1	2012	2012	2	2.35E-07	2.35E-07	2.11323E-06
2	2012	2012.083333	1	2.03E-07	2.03E-07	nan
3	2012	2012.166667	3	1.22E-06	1.22E-06	3.03627E-06
4	2012	2012.25	5	2.69E-07	2.69E-07	3.3389E-07
5	2012	2012.333333	1	5.34E-07	5.34E-07	nan
6	2012	2012.416667	2	1.72E-06	1.72E-06	1.54383E-05
7	2012	2012.5	2	1.02E-07	1.02E-07	9.19857E-07
8	2012	2012.583333	2	2.04E-05	2.04E-05	0.000183349
9	2012	2012.666667	3	7.30E-06	7.30E-06	1.81438E-05
10	2012	2012.75	3	1.86E-07	1.86E-07	4.62574E-07
11	2012	2012.833333	3	2.76E-07	2.76E-07	6.86494E-07
12	2012	2012.916667	2	2.38E-08	2.38E-08	2.13459E-07
1	2013	2013	2	5.60E-08	5.60E-08	5.0301E-07
2	2013	2013.083333	1	1.33E-08	1.33E-08	nan
3	2013	2013.166667	4	7.71E-07	7.71E-07	1.22753E-06
4	2013	2013.25	2	6.21E-07	6.21E-07	5.58366E-06
5	2013	2013.333333	2	3.82E-07	3.82E-07	3.43451E-06
6	2013	2013.416667	4	3.14E-07	3.14E-07	4.99657E-07
7	2013	2013.5	1	1.34E-07	1.34E-07	nan
8	2013	2013.583333	1	9.70E-08	9.70E-08	nan
9	2013	2013.666667	2	1.27E-07	1.27E-07	1.1406E-06
10	2013	2013.75	4	5.98E-07	5.98E-07	9.51518E-07
11	2013	2013.833333	1	4.12E-09	4.12E-09	nan
12	2013	2013.916667	1	1.44E-07	1.44E-07	nan
1	2014	2014	6	7.61E-07	7.61E-07	7.98972E-07
2	2014	2014.083333	1	1.30E-07	1.30E-07	nan
3	2014	2014.166667	3	1.70E-07	1.70E-07	4.2165E-07
4	2014	2014.25	5	6.79E-06	6.79E-06	8.43607E-06
5	2014	2014.333333	1	1.41E-06	1.41E-06	nan
6	2014	2014.416667	2	2.51E-06	2.51E-06	2.25672E-05
7	2014	2014.5	6	7.89E-07	7.89E-07	8.28446E-07
8	2014	2014.583333	2	8.90E-07	8.90E-07	7.99612E-06
9	2014	2014.666667	1	3.35E-08	3.35E-08	nan
10	2014	2014.75	2	3.31E-07	3.31E-07	2.97659E-06
11	2014	2014.833333	3	8.71E-08	8.71E-08	2.16324E-07
12	2014	2014.916667	1	1.65E-07	1.65E-07	nan
1	2015	2015	1	5.49E-08	5.49E-08	nan
2	2015	2015.083333	2	7.78E-06	7.78E-06	6.99227E-05
3	2015	2015.166667	1	3.29E-09	3.29E-09	nan
4	2015	2015.25	2	8.58E-08	8.58E-08	7.71105E-07
5	2015	2015.333333	1	2.21E-08	2.21E-08	nan
6	2015	2015.416667	0	0	0	nan
7	2015	2015.5	2	2.89E-06	2.89E-06	2.59346E-05
8	2015	2015.583333	4	3.18E-07	3.18E-07	5.06407E-07
9	2015	2015.666667	3	3.81E-07	3.81E-07	9.45358E-07
10	2015	2015.75	1	2.92E-08	2.92E-08	nan
11	2015	2015.833333	1	2.04E-09	2.04E-09	nan
12	2015	2015.916667	4	2.01E-07	2.01E-07	3.20049E-07
1	2016	2016	1	4.30E-07	4.30E-07	nan
2	2016	2016.083333	2	3.81E-08	3.81E-08	3.42602E-07
3	2016	2016.166667	2	1.16E-07	1.16E-07	1.04312E-06
4	2016	2016.25	4	1.23E-07	1.23E-07	1.94928E-07
5	2016	2016.333333	1	9.89E-08	9.89E-08	nan
6	2016	2016.416667	6	7.67E-07	7.67E-07	8.04679E-07
7	2016	2016.5	3	3.37E-06	3.37E-06	8.36179E-06
8	2016	2016.583333	3	2.81E-06	2.81E-06	6.97028E-06
9	2016	2016.666667	5	3.84E-06	3.84E-06	4.76327E-06
10	2016	2016.75	1	2.23E-09	2.23E-09	nan
11	2016	2016.833333	7	2.32E-06	2.32E-06	2.14818E-06
12	2016	2016.916667	3	1.61E-06	1.61E-06	4.00198E-06
1	2017	2017	6	6.07E-07	6.07E-07	6.36821E-07
2	2017	2017.083333	3	5.94E-07	5.94E-07	1.47632E-06
3	2017	2017.166667	7	8.59E-07	8.59E-07	7.94011E-07
4	2017	2017.25	5	1.78E-06	1.78E-06	2.21358E-06
5	2017	2017.333333	1	3.58E-08	3.58E-08	nan
6	2017	2017.416667	3	4.70E-07	4.70E-07	1.1678E-06
7	2017	2017.5	5	1.19E-06	1.19E-06	1.47301E-06
8	2017	2017.583333	1	1.92E-07	1.92E-07	nan
9	2017	2017.666667	4	8.85E-07	8.85E-07	1.40827E-06
10	2017	2017.75	0	0	0	nan
11	2017	2017.833333	3	1.46E-06	1.46E-06	3.62861E-06
12	2017	2017.916667	5	7.81E-08	7.81E-08	9.7011E-08
1	2018	2018	0	0	0	nan
2	2018	2018.083333	3	2.06E-07	2.06E-07	5.12779E-07
3	2018	2018.166667	4	1.71E-06	1.71E-06	2.71443E-06
4	2018	2018.25	5	1.73E-07	1.73E-07	2.14767E-07
5	2018	2018.333333	3	1.26E-06	1.26E-06	3.13802E-06
6	2018	2018.416667	1	1.34E-07	1.34E-07	nan
7	2018	2018.5	7	1.03E-06	1.03E-06	9.55121E-07
8	2018	2018.583333	4	5.77E-07	5.77E-07	9.1807E-07
9	2018	2018.666667	9	1.31E-06	1.31E-06	1.00783E-06
10	2018	2018.75	6	1.18E-07	1.18E-07	1.24229E-07
11	2018	2018.833333	2	6.31E-08	6.31E-08	5.66748E-07
12	2018	2018.916667	4	3.25E-07	3.25E-07	5.17668E-07
1	2019	2019	4	4.18E-07	4.18E-07	6.6514E-07
2	2019	2019.083333	3	3.46E-07	3.46E-07	8.60735E-07
3	2019	2019.166667	8	2.38E-06	2.38E-06	1.98943E-06
4	2019	2019.25	11	9.46E-06	9.46E-06	6.35449E-06
5	2019	2019.333333	2	1.99E-06	1.99E-06	1.78939E-05
6	2019	2019.416667	5	2.05E-06	2.05E-06	2.54425E-06
7	2019	2019.5	11	1.23E-06	1.23E-06	8.28714E-07
8	2019	2019.583333	11	6.69E-07	6.69E-07	4.49187E-07
9	2019	2019.666667	8	5.20E-07	5.20E-07	4.34776E-07
10	2019	2019.75	13	9.13E-07	9.13E-07	5.51655E-07
11	2019	2019.833333	7	6.15E-06	6.15E-06	5.69019E-06
12	2019	2019.916667	10	5.86E-07	5.86E-07	4.19217E-07

}{\data}

\begin{tikzpicture}
\begin{axis} [
    width=2.9in,
    height=2.0in, 
    xlabel={time},
    xlabel near ticks,
    xticklabels={ 2010, 2012, 2010, 2012, 2014, 2016, 2018, 2020},
    x tick label/.append style={
        /pgf/number format/.cd,
        scaled x ticks = false,
        set thousands separator={},
        fixed
        },
    ylabel style={align=center},
    ylabel={avg. max normalized\\meme-frequency},
    ylabel near ticks
]

\addplot [blue] table [x=date, y=avg_peak_freq] {\data};
\addplot [name path=upper,draw=none] table[x=date ,y expr=\thisrow{avg_peak_freq}+\thisrow{95_confidence}] {\data};
\addplot [name path=lower,draw=none] table[x=date ,y expr=\thisrow{avg_peak_freq}-\thisrow{95_confidence}] {\data};
\addplot [fill=blue!20] fill between[of=upper and lower];

\end{axis}
\end{tikzpicture}
    \vspace{-6mm}
    \caption{Average maximum normalized meme-frequency occurring in each month. Shaded regions indicate 95\% confidence intervals. Maximum frequency of memes is consistent when growth of Reddit is controlled for, because limited attention bounds each meme.\tw{candidate for deletion} \rk{I can't get the y-label to line break, somebody HELP!} \rk{WARNING!!!!!!! don't resurrect this plot without updating the data first!!!!!}}
    \label{fig:avg_meme_peak}
\end{figure}
}

\section{Conclusions}



The three research questions in the present work coalesce into an emerging theory of meme dynamics in online social bulletin boards like Reddit. Taking an ecological perspective, we have shown that although Reddit as a whole continues to grow, the fraction of Reddit devoted to text memes is consistent. This reinforces the notion of a meme ecology, which, like the economic and epidemiological perspectives of information diffusion, has consequences that derive from limited user attention. 

We have also shown that the diversity of memes across communities is decreasing slightly, even as the number of communities continues to grow. This represents yet another consequence of the ecological perspective: as social media communities continue to Balkanize, \ie, split into narrow and self-referential communities, the number of shared references to memes appears to be decreasing.

The competition among memes is also evident in the increasing turnover of beachhead communities. As memes fight for user and subreddit attention, cycling in and out of collective attention faster, large and general communities are unable to consistently integrate new memes before other smaller communities. Instead, subreddits may rise to the top if they happen to introduce the meme-of-the-moment, only to quickly fade into obscurity.

Finally, we have shown that meme lifespans have decreased significantly. Yet, unlike similar work on general n-grams, hashtags, citations, etc., we do not find that all aspects of meme dynamics are accelerating, and even the accelerations are bounded by the effect of increasing competition for limited user attention. The patterns of relative meme growth, even surrounding a meme's peak, remain relatively unchanged by the expansion of Reddit. And although the absolute number of memes active on Reddit has increased over time, this growth has been outpaced by the growth of Reddit itself. 

It is important to recognize the limitations of this study. Our evaluation is principally limited by the KnowYourMeme dataset. Because the popularity metric is accumulative on KnowYourMeme, it is likely that newer memes, which have not yet accumulated their full/final popularity, are less likely to appear in our dataset. Our focus on the textual artifacts of memes may also limit the generalizability of our findings. Much of the online meme culture, especially recently, is conveyed through visual depictions of memes~\cite{theisen2020automatic}. We necessarily assume that textual representations of visual memes follow proportionately, but we are unable to validate this claim in the current work. Future work should endeavor to confirm these findings on visual imagery. Finally, the present work used a strict text-matching algorithm to identify memes. Additional follow-up work is needed to better understand and trace how these memes evolve and how the shape of a meme's popularity and evolution affects its lifespan and reach.


\section*{Acknowledgements}
We would like to thank Satyaki Sikdar for his help preparing this manuscript. This work is funded by the US Army Research Office (W911NF-17-1-0448) and the US Defense Advanced Research Projects Agency (DARPA W911NF-17-C-0094).

\bibliographystyle{plain}
\bibliography{ms.bib}

\end{document}